# CONCEPTUAL DESIGN OF 20 T DIPOLES FOR HIGH-ENERGY LHC

L. Rossi, E. Todesco, CERN, Geneva, Switzerland

*Abstract*

Availability of 20 T operational field dipole magnets would open the way for a 16.5 TeV beam energy accelerator in the LHC tunnel. Here we discuss the main issues related to the magnet design of this extremely challenging dipole: main constraints, superconductor choice, coil lay-out, iron, forces and stresses, and field quality. A tentative cost estimate is also given. The present technology, based on Nb-Ti and now near to be extended to $Nb_3Sn$ superconductor, would allow reaching 15 T operational field. To reach 20 T, HTS conductors capable to carry 400 A/mm$^2$ at 15-20 T under transverse stress of 150-200 MPa are an essential element.

## INTRODUCTION

The LHC main dipoles [1] are today running at 4.15 T, i.e., about 0.1 T less than Tevatron dipoles [2], which are based on the same Nb-Ti superconductor, and were built more than thirty years ago. After the consolidation of the splices in the magnet interconnects [3], the LHC main dipoles will be in conditions for reaching the design field value of 8.3 T. This will not happen before 2013. Presenting today a study for a 20 T dipole for a new machine to be installed in the LHC tunnel may seem, and actually is, a huge leap.

Indeed, the timeline of development of superconducting magnets for accelerators is long: for Nb-Ti based magnets, which is a well assessed technology for accelerators, the experience gained in the construction of several accelerators shows that five years are needed from day-zero, when aperture and field are decided, to installation and commissioning. For more performing and complex technology, like the one based on $Nb_3Sn$ technology, the time is longer: the vigorous LARP program [4] took more than five years to successfully build a 3.4-m-long model quadrupole [5], which is a bare quadrupole with no cryostat and other integration features and, as magnet, only partly satisfies the requirements needed for installation in the LHC. Whereas for $Nb_3Sn$ the conductor with the many – although not all – required properties is today available, in the case of high temperature superconductors (HTS), substantial improvement of the basic performance of the conductor itself is needed, both in terms of current density and strain degradation [6]. This implies even much longer times. Therefore for making credible the High Energy LHC (HE-LHC) as one of the options for CERN after the LHC, i.e., around 2030, it is necessary starting now to explore the main issues related to the magnet design, and to drive the R&D in the needed superconductors.

The maximum field reached in an accelerator-type dipole is around 14 T at 4.5 K [7], using $Nb_3Sn$ conductor, in an aperture similar to the HE-LHC requirements (40 mm). It should be noted that in more than 10 years no dramatic improvement happened after the 13.5 T at 2 K in a 50 mm bore reached in 1997 by the D20 dipole [8]. Due to the shape of the critical surface, the maximum field attainable with $Nb_3Sn$ accelerator magnets is around 18 T. May be 19 T could be reached with an optimized superconductor lay-out. Taking 18 T as solid figure, for the HE-LHC this gives ~15 T operating field after imposing the 20% margin, that at this stage we assume as reasonably needed for a series production of more than 1000 magnets. Of course this assumption can be challenged: however the experience of past accelerators (see Table 1) shows that a solid margin in the design is needed to compensate inevitable non-homogeneity of about 10% in performance.

Table 1: Operational dipole field, current and operational margin in high energy physics accelerators

|  | Operational field (T) | Operational current (kA) | Operational margin (%) |
|---|---|---|---|
| Tevatron | 4.4 | 4.3 | ~26% |
| HERA | 4.7 | 5.0 | ~31% |
| RHIC | 3.5 | 5.5 | ~33% |
| *SSC* | *6.7* | *6.6* | *~15%* |
| LHC | 8.3 | 11.8 | ~16% |

*SSC was cancelled in 1993 after 10 years of R&D and prototypes*, HERA operation field was increased at 5.5 T (limiting margin reduction by lowering temperature down to 3.9 K) in 1998 [9].

Superconducting cables based on HTS are able to withstand fields larger than 15 T: they have been successfully used in high-field solenoids [6] but not in accelerator dipoles.

From the point of view of magnet design, a 20 T dipole for the LHC poses two big challenges: (i) obtain such a high field with a compact coil, and shield it with enough iron without exceeding the transverse dimensions imposed by the LHC tunnel; (ii) manage the stresses induced by electromagnetic forces to avoid degradation of the conductor.

$Nb_3Sn$ is more than a factor five more expensive than Nb-Ti. Similarly, HTS is another factor 3-5 more expensive than $Nb_3Sn$. It is unlikely that the large difference in price between the three superconductors will disappear, even in the time scale of a production of the HE-LHC magnets (2025, i.e., 15 years from now). For this reason, a hybrid coil is required to minimize the cost of the conductor, which is a large fraction of the whole project. The construction of a hybrid coil poses the third difficult challenge: each material needs different heat treatments, needs different approach to stability and mechanical structure, and there is very little experience in building hybrid magnets for accelerators [10].

The proposal of an 'LHC energy upgrade' dates back to early 2000 [11] and a lay out for a 24 T (short sample, i.e., with no operational margin) hybrid magnet was proposed in 2005 [12]. The new name HE-LHC looks more appropriate, since here we are talking about replacing at least all the LHC magnets, i.e., building practically a new machine, since many other systems will have to be upgraded or modified [13]. However, the main infrastructures of the tunnel (the 27 km of LHC machine and the 6 km of injection transfer lines with many of the technical services) would be kept or just consolidated, giving a major advantage w.r.t. other projects needing new tunnels and new infrastructure.

## CONSTRAINTS AND ASSUMPTIONS

### Aperture

The LHC accelerates particles from 450 GeV to 7 TeV, i.e. a factor 15.6 [14], to be compared to a factor 6 in Tevatron [2] and 25 in HERA [15]. Acceleration from 450 GeV to 16.5 TeV in the HE-LHC would imply more than a factor 30 of acceleration. Injection at 1.2 TeV brings the energy increase to a factor 14, and allows reducing the aperture of the machine, which is a critical parameter both in terms of cost and transverse size. About 1/3 of the 56 mm aperture of LHC main magnets is used for beam tubes, beam screen and clearance, while the rest are available for the beam. The beam size scales with the square root of the inverse of the energy. Increasing injection energy from 0.45 to 1.2 TeV, the aperture available for the beam can be reduced by ~40%: therefore the total aperture of the main magnets can go to ~40 mm. Certainly a study and subsequent optimization can indicate how much the aperture can be further reduced, below 40 mm. However this is a reasonable guess, especially for 15-m-long and curved dipoles. *With an injection at 1.2 TeV an aperture of 40 mm is considered.*

### Magnet size and current density

The 3.8 m diameter of the LEP tunnel where the LHC is located is a strong constraint on the magnet transverse size, despite the space saving due to the twin design. In the LHC, the cold mass has a diameter of 570 mm. This size $d_{cm}$ is given by

$$d_{cm}= d_b+2(r+c_t+s_t+i_t+S_t)$$

where $d_b$ = 192 mm is the beam separation, $r$=28 mm is the aperture radius, $c_t$=31 mm is the coil thickness, $s_t$=40 mm is the structure (collar) thickness, $i_t$=80 mm the iron thickness, and $S_t$=10 mm the shell thickness (see the sketch shown in Fig. 1).
The field in a dipole is proportional to the coil thickness and to the current density. For a 60° sector coil one has

$$B[T]=0.00069\, j_o\, [A/mm^2]\, c_t\, [mm]$$

An analysis of the relation coil thickness vs. operational field in accelerator magnets shows that they are not so far from the line corresponding to a overall current density of 400 A/mm$^2$, i.e., as in the LHC (see Fig. 2).

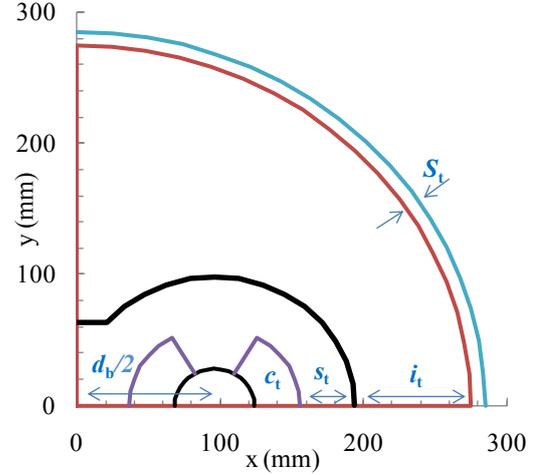

Figure 1: Schematic cross-section of the LHC dipole, one quarter shown.

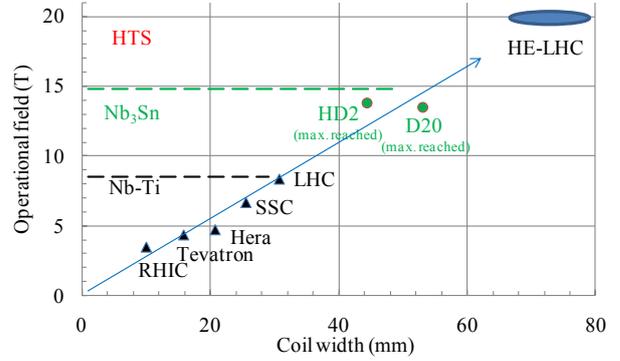

Figure 2: Operational field versus coil width in Nb-Ti accelerator magnets. For Nb$_3$Sn models the maximum reached field is given. The straight line fit has a slope consistent with $j_0$= 400 A/mm$^2$.

Taking for the HE-LHC dipole the same current density as in the LHC as a first guess, the coil thickness should be increased by a factor 2.5 from 30 to 75 mm, and the iron needed to shield scales with aperture and field from 80 to 130 mm. This gives a cold mass diameter ~300 mm larger than the LHC dipoles: this is close to the upper limit fixed by the requirements for installation and transport. This first estimate suggests that *the current density cannot be much lower than in the LHC coil, i.e. 350-400 A/mm$^2$.*

### Cost

The cost of the conductor in the LHC main dipoles is approximately one third of the cost of the magnet (300 kCHF out of 1 MCHF). A coil with a thickness of 75 mm and an aperture of 40 mm has 3.2 times the surface of the LHC coil. The 8 T operational field is the upper limit of what can be reached with Nb-Ti. Nb$_3$Sn allows reaching operational fields in the range of ~15 T, as foreseen for the High Luminosity LHC on the 2020 horizon, but today it is at least 5 times more expensive than Nb-Ti. A coil made of Nb$_3$Sn would cost about 3.2×5=16 times the LHC dipole coil, i.e., about 5 MCHF per magnet, and would not reach the 20 T. High temperature superconductors are 5 times more expensive than Nb$_3$Sn (with large variations): a coil made only of HTS would cost the stellar price of 3.2×5×5, i.e. 80 times an LHC coil (24 MCHF per magnet!). We assume that even in the time scale of the HE-LHC (i.e., 20 years from today) the large difference in price will not disappear. Therefore, following what is done in high field solenoids, *one has to build a hybrid coil, where cheaper superconductors are used in the lower field regions.*

*Margin*

We assume that the magnets will operate at 80% from the critical surface, i.e. a 20% operational margin. This may appear a rather conservative assumption: LHC magnets have a 14% operational margin (see Table 1). However, they still have to reach, in the machine, the operational field (and most probably a long training is needed [16]). With Nb$_3$Sn, there is not enough experience to firmly establish the needed margin, which is a rather controversial parameter, and could range between 10% and 20%: here we take a conservative estimate.

*Stress*

Both Nb$_3$Sn and HTS materials can undergo a severe degradation due to strain [17]. For this reason, the level of stress in the coil due to electromagnetic forces is a critical issue. In the LHC, the coil stress due to electromagnetic forces is of the order of 70 MPa [18]. Since the force scales with the field times the current density, going to 20 T with the same current density brings stresses to 150-200 MPa, which is the range where considerable degradation of Nb$_3$Sn starts (actually for certain type of Nb$_3$Sn serious degradation occurs even above 120 MPa). Therefore, *the stress constraints prevent from using higher current densities than what we have in the LHC dipoles.*

## THE HYBRID COIL LAY-OUT

In the lower field region, the first 8 T are obtained with Nb-Ti conductor as in the LHC coils. We assume an overall current density (i.e., the current density of the coil, including voids and insulation, but not copper wedges) of 380 A/mm$^2$. This corresponds to a filling factor of 0.35 (i.e., 35% of the cross section of the coil is made of superconductor, and the rest is stabilizer, passive elements, void and insulation). For comparison, the LHC dipole inner cable has a 0.33 filling factor with a copper-superconductor ratio of 1.65. Using these parameters, one can reach 8 T in the Nb-Ti coils, with a 20% operational margin (see Fig. 3), similarly to the LHC case.

For Nb$_3$Sn we assume a rather conservative estimate for the superconductor current density of 2500 A/mm$^2$ at 12 T and 4.2 K, or 1250 A/mm$^2$ at 15 T and 4.2 K. This corresponds to 480 A/mm$^2$ at 16 T and 1.9 K of overall current density, with a filling factor of 0.3. These values allow reaching 13 T with a 20% operational margin (see Fig. 3). To further reduce the quantity of HTS, we suggest using a lower current density of ~200 A/mm$^2$ in the field region beyond 13 T. This makes the coil larger, but allows reaching 15 T (see Fig. 3, lower loadline), besides helping to reduce mechanical stresses.

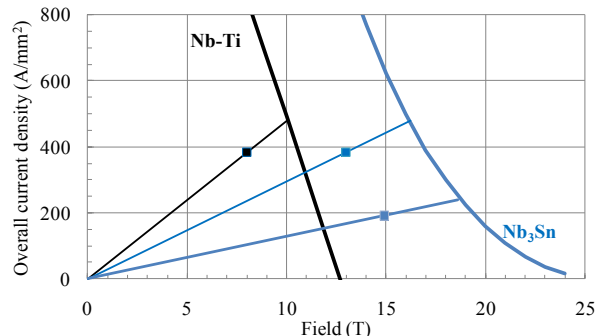

Figure 3: Overall current density in Nb-Ti and Nb$_3$Sn (curves), loadlines (straight lines) and operational points.

The last 5 T must be provided by HTS: a further reduction of a factor two in Nb$_3$Sn current density would give about 100 A/mm$^2$, and to gain another 2 T one would have to add 40-mm-thick coil, that would probably increase the transverse size beyond our constraints.

Among the HTS, Bi-2212 has the advantage of being available in form of round wires, but has low engineering current density and large strain degradation. The alternative is YBCO, which has a much lower degradation, higher current density but no round wire. Independently of this choice, we assume to have a cable operating with 380 A/mm$^2$ overall current density. This is about twice of what can be obtained today for Bi-2212, however there is consensus that with a vigorous R&D this value can be obtained in industrial scale, very much like Nb$_3$Sn that, by means of the US-DOE program [19] has more than doubled its current density in 10 years.

A cross-section with 11 blocks drawn according to the above guidelines is shown in Fig. 4. The two outer blocks, where the field reaches 8 T, are made with Nb-Ti. Then one has four blocks with Nb$_3$Sn, three blocks with Nb$_3$Sn at half current density, and two blocks with HTS. With this highly optimized cross-section the fraction of HTS is about 1/6, almost 1/3 is Nb-Ti, and more than half is Nb$_3$Sn (see Table 2). We use the cable geometry of HD2, with 28×2 0.8 mm strands, 22.2 mm width and 1.62 mm thickness, and with an insulation of 0.11 mm. A total of 150 turns are needed. Operational current for 20 T, with the iron described in next section, is 6.9 kA in the low density Nb$_3$Sn region and 13.8 kA elsewhere.

With respect to the pioneering work presented in [12] (see Fig. 5), where the current density was set at 800 A/mm$^2$, based on an optimistic guess of the progress in the Nb$_3$Sn and HTS development, and on the principle of stress management that removes one constraint, here we are at half of the current density. This doubles the quantity of superconductor, see Table 2. Indeed, thanks to the optimization of the grading and to the use of Nb-Ti, we manage to end up with 25% less HTS conductor w.r.t. [12]. With respect to the layout shown in Fig. 5, our proposal leaves no space for a support structure between the blocks (see Fig. 4): this aspect should be carefully considered and could be critical. We consider a two-in-one geometry as in the LHC; the common coil option [20,21] should be also investigated.

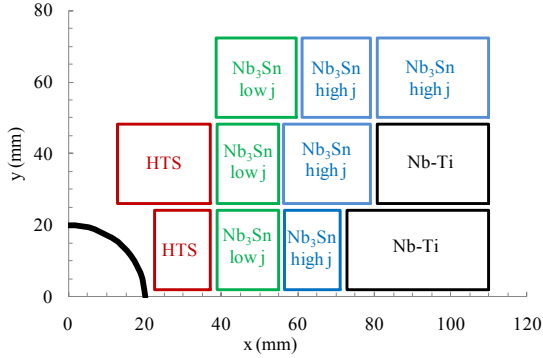

Figure 4: Block lay out of the coil (one quarter of one aperture only is shown).

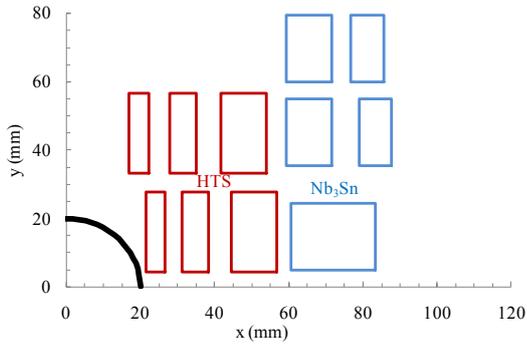

Figure 5: Block lay out of the coil proposed in [11] (one quarter of one aperture shown).

Table 2: Coil cross section (for one aperture) for layouts shown in Fig. 4 and 5.

|  | This work | | Ref [13] | |
| --- | --- | --- | --- | --- |
|  | Surface (cm$^2$) | % | Surface (cm$^2$) | % |
| Nb-Ti | 59 | 27% | - | - |
| Nb$_3$Sn | 122 | 57% | 50 | 52% |
| HTS | 35 | 16% | 46 | 48% |
| Total | 216 | | 96 | |

## THE IRON

We use a 120 mm thick iron, placing it as close as possible to the coil. In this structure, collars are replaced by spacers, and the forces are kept by the iron-shell part. Self-supporting collars would need additional space.

The peak field in the coil blocks, in presence of iron with an external diameter of 800 mm and at operational field of 20 T in the bore, is shown in Fig. 6. The iron is placed at a larger distance in the inner part of the coil (i.e., the part towards the centre of the magnet) to reduce the influence of one aperture on the other (see Fig. 7). In fact the two-in-one structure induces higher peak field in the side of the coil which points at the centre of the magnet, and the iron can be used to partially compensate this effect. This cross-talk also requires to have some space between the coils of the two apertures, and brings the beam separation from the 192 mm of the LHC to 300 mm. Eliminating this space one could save 100 mm in the magnet size, but the margin would be largely reduced. The iron contributes to about 7% of field for a fixed current. Computations were done with ROXIE [22].

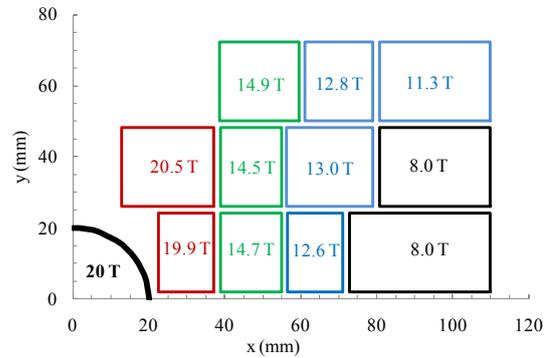

Figure 6: Peak field in the blocks at 20 T field.

Iron is largely saturated at 20 T operational field (see Fig. 7). The fringe field at 200 mm from the cold mass is 20 mT, which is within the specification for the LHC tunnel (50 mT). The iron thickness should be reduced in a more refined design, since for example we have not yet considered the thickness of the restraining cylinder. Since the 800 mm here given for the iron yoke is considered the maximum allowable diameter (and maybe even beyond!) to stay in a cryostat compatible with the LHC tunnel, this means that the 50 mT threshold should be either reached or passed. A solution may be in considering anti-coils to shield the field demagnetizing the outer iron: this solution is routinely used in MRI solenoids, but may be very difficult in dipoles. In alternative, a different lay-out of the cryostat and cryogenic system must be explored: for example reconsidering integrating the cryolines inside the magnet cryostat, like in the original LHC design [23] and in all other accelerators. This would allow larger cryostat and larger cold mass. Clearly this is a critical point to be addressed with a deep and wide investigation. A summary of the main parameters is given in Table 3. The very large stored energy (13 times the LHC dipoles) represents a big challenge for magnet protection.

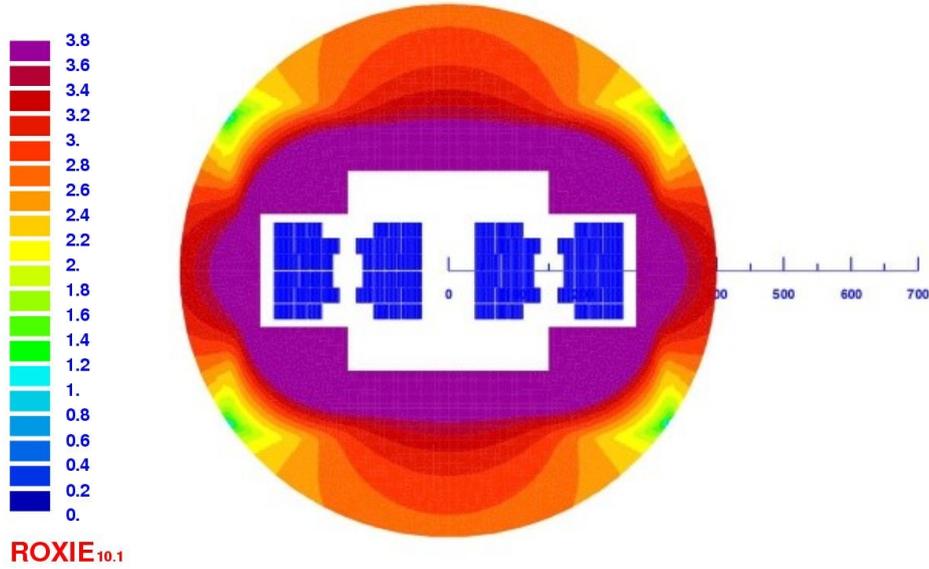

Figure 7: Cross-section of the magnet (coil, structure and yoke), showing field in the iron (color code is in tesla). The horizontal axis is in mm.

Table 3: Main parameters of the HE-LHC and LHC dipole

|  |  | HE-LHC | LHC |
|---|---|---|---|
| Operational field | (T) | 20.0 | 8.3 |
| Operational current | (kA) | 13.8/6.9 | 11.8 |
| Operational margin | (%) | 20 | 14 |
| Magnetic lenght | (m) | 14.3 | 14.3 |
| Total stored energy | (MJ) | 100 | 7.0 |
| Distance between beams | (mm) | 300 | 194 |
| Total number of turns | (adim) | 150 | 40 |
| Cable width (bare) | (mm) | 22.2 | 15.1 |
| Cable thickness (bare) | (mm) | 1.62 | ~1.9/1.5 |
| Insulation thickness | (mm) | 0.11 | 0.12 |
| Maximum coil thickness | (mm) | 97.3 | 31 |
| Coil height | (mm) | 72.2 | - |
| Cold mass diameter | (mm) | 800 | 570 |

## FIELD QUALITY

The proposed dipole layout has a ratio between the coil width and the aperture radius of ~4 (see Fig. 8). This ratio is a relevant parameter for field quality: the larger it is, the lowest are the high order multipoles, since a good part of the coil is 'far' from the beam, and therefore contributes only to the main component and not to the high order harmonics. This is why the multipole optimization is easier w.r.t. accelerator magnets which have a much lower ratio (see Fig. 8). In our case, the cross-section shown in Fig. 7 has all field harmonics within 2 units without the need of any copper wedge! The horizontal position of the three layers provides three free parameters which are enough to minimize all harmonics. The large saturation of the iron should pose no particular problems for operation, and the impact on $b_2$ could be corrected through quadrupoles.

On the other hand, persistent current will create large components at injection. The filaments in $Nb_3Sn$ and HTS are at least a factor seven larger w.r.t. Nb-Ti, and since these components scale with the filament size, they will be much more relevant than in the present LHC dipoles.

This could induce a large change of $b_3$ during the ramp, to be corrected through spool pieces. Surprisingly $Nb_3Sn$ has neither decay nor snapback [24]: this feature, which is not yet understood, would greatly ease operation. Cable effects needs have not yet been studied: interstrand resistance is more difficult to control than in Nb-Ti.

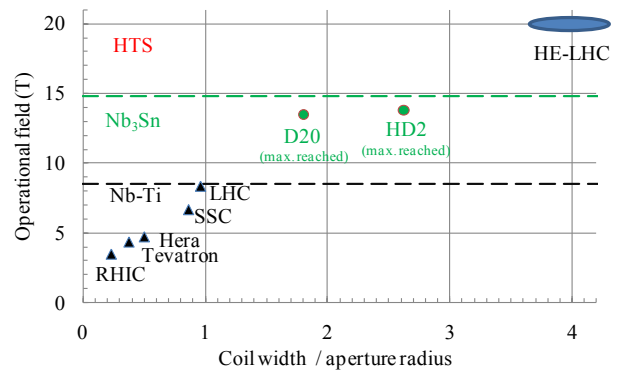

Figure 8: Operational field versus ratio coil width/aperture radius in Nb-Ti accelerator magnets. For $Nb_3Sn$ models the maximum reached field is given.

## STRESS

The use of a low current density has the drawback of giving a less compact coil, but the advantage of giving less stress. With respect to the 800 A/mm$^2$ used in [12], with half current density we manage to keep stresses at a lower level. Here we give a first estimate [25] based on a coil where the block are completely glued one to the others, without structure around, and not pre-stressed (see Fig. 9). In this zero-order case one sees that the higher stresses of about 220 MPa are in the Nb-Ti region, that in $Nb_3Sn$ stress is below 180 MPa and even lower in the HTS.

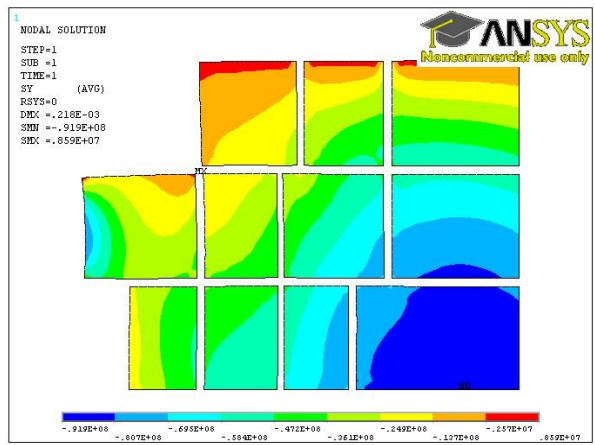

Figure 9: Estimate of horizontal stress at operational current, glued case without preloading.

Indeed, one has to take into account that the coil needs to be pre-stressed with a horizontal load which usually is 80%-100% of the maximum stress in operational conditions. Therefore, if we stay on the lower side, during assembly a uniform horizontal stress of 180 MPa should be envisaged. This is tolerable for Nb-Ti and just acceptable for $Nb_3Sn$. It is well beyond what Bi-2122 can withstand, but we can hope it can (or it will be) tolerated by YBCO based superconductors which are intrinsically quite robust, thanks to the steel substrate. However, compression stress limit in HTS needs to be addressed by a proper R&D program.

## COST

As we are talking about prices in 2025, the cost estimate is a difficult and acrobatic exercise. Indeed, this is an essential ingredient of the story! To avoid writing something that becomes outdated or simply wrong tomorrow, one has to clearly state the hypothesis of our estimate. For the conductor, we consider 200 $/kg for Nb-Ti, that is the present cost. The $Nb_3Sn$ price ranges today between 1000 and 1300 $/kg: we assume a price of 800 $/kg., i.e., a 20% improvement w. r. t. the cheapest price. For HTS we assume 3000 $/kg, which is the lower edge of today price, but for a material not reaching our required performances. Under these assumptions, the total cost of the conductor per magnet is 3.8 M$, half of which is for the last 5 T with HTS (see Table 4).

On the top of this, the manufacturing cost has to be added. For the LHC we had, as rough figures, 300 kCHF of components, 300 kCHF of conductor, 300 kCHF of assembly, and 100 kCHF of cryostat, testing, etc. The main difference for a 20 T magnet would be the coil manufacturing (100 kCHF out of 300 kCHF for the LHC). Doubling this component and keeping the same value for the other items, we would reach 800 kCHF of assembly, components and cryostats. This gives a final cost of 4.6 M$ per magnet (at the moment, 1$~1CHF~0.77 euro). At this level of a very preliminary budgetary estimate, choosing dollars, euros or Swiss francs (and guessing the exchange rate in 15 years …) is within the error of our estimate. Having 1200 magnets, the total cost of magnet would be around 5500 M$.

Table 4: Estimate of the cost of the conductor for a 14.3 m coil length two-in-one dipole.

|  | ($/kg) | $m^3$ | Kg | M$ | % | Field (T) |
|---|---|---|---|---|---|---|
| Nb-Ti | 200 | 0.12 | 960 | 0.19 | 5% | 8 |
| $Nb_3Sn$ - h | 800 | 0.16 | 1300 | 1.0 | 28% | 13 |
| $Nb_3Sn$ - l | 800 | 0.10 | 850 | 0.7 | 18% | 15 |
| HTS | 3000 | 0.07 | 620 | 1.9 | 49% | 20 |
| Total |  | 0.45 | 3730 | 3.8 |  |  |

This may seem, and it is, a very high cost. Indeed, it is interesting to compare it with what could be done tomorrow with present technology: an accelerator with dipoles at 8 T, whose arcs are 2.5 longer than in the LHC. This machine would need 3000 LHC magnets for a total cost of 3000 M$. On the top of this, one should add the cost of the 65-km-long tunnel which can be estimated between 1000 and 2000 M$: this brings the total in the same range. A larger machine would then need new cryogenics, and infrastructures, whereas the HE-LHC would need an additional injector, the cost of the second being probably lower. Finally one would probably need new infrastructures for experiments.

A clear drawback shown by this preliminary analysis is that the cost of this project would be largely dominated by two components: the $Nb_3Sn$ cable and the HTS cable, sharing each of them about one third. This is a risk for a large project, taking into account that at the moment very few producers are present on this market: for instance, the $Nb_3Sn$ cable of the LARP, which is leading the high field magnet research, all comes from the same manufacturer.

Lowering the target from 20 T to 15 T would considerably reduce the price, possibly by 30%. Nevertheless, given the long timeline we are considering, we believe that there are considerable margins for HTS improvement, also in term of cost reduction. A recent DOE program on Bi-2212 goes in this direction. As a matter of facts, the development of HTS material has been mainly driven by applications that are far away from high energy physics, and a different strategy could lead to relevant improvements in the direction useful for the HE-LHC.

## CONCLUSION

In this paper we explored the possibility of having 20 T operational field dipole magnets in the LHC tunnel. Other important magnets, like main quadrupoles have not been studied and are shortly addressed in another paper [26].

Main constraints are (i) the transverse size of the magnet, limited by the tunnel, (ii) the stress in the coil induced by electromagnetic forces, and (iii) the cost. All these constraints call for a design based on hybrid coils that allows using cheaper conductor in the lower field regions. A hybrid layout, based on Nb-Ti, $Nb_3Sn$ and HTS, that meets all basic requirements (including 20% field margin) is then proposed and examined. With respect to previous work [12] we reduced the overall current density from 800 A/mm$^2$ to 400 A/mm$^2$, plus a special region at 200 A/mm$^2$ to allow reaching 15 T with $Nb_3Sn$. This gives lower stresses, at the limit of what is manageable for $Nb_3Sn$, and allows using the HTS only in the 15 to 20 T field regions. This layout requires an HTS cable based on round wire, capable of carrying 400 A/mm$^2$ overall current densities at 15-25 T under 180 MPa compressive stresses, not yet available today. The main targets for future R&D should be directed toward the 13-15 T region, where $Nb_3Sn$ good results on small coils need to be consolidated and oriented toward accelerator quality, and toward a basic improvement of HTS in term of critical current, stress tolerance and suitability to be assembled large current compact cable. The R&D on HTS is critical, also in term of time, if the goal of 20 T for 2030 has to remain credible; if in a few years new results will not be available, the HE-LHC should be reduced its target to 15 T (may be 16 T with a suitably optimized design) for the main dipole field.

## ACKNOWLEDGEMENTS

We wish to thank B. Auchmann and S. Russenschuck for help building the electromagnetic model and in the energy computation, and A. Milanese for building the mechanical model and estimating the stresses.

## REFERENCES

[1] L. Rossi, "Superconducting magnets for the LHC main lattice", IEEE Trans. Appl. Supercond. 14 (2004) 153.
[2] R. Hanft, et al., "Magnetic field properties of Fermilab Energy Saver Dipoles", TM-1182, 1630, 03/1983.
[3] F. Bertinelli, International Particle Accelerator Conference (2011), to be published.
[4] G. L. Sabbi, these proceedings.
[5] G. Ambrosio, et al., "Final development and test preparation of the first 3.7 m long Nb3Sn quadrupole by LARP", IEEE. Trans. Appl. Supercond. 20 (2010) 283.
[6] J. Schwartz, these proceedings.
[7] P. Ferracin, et al., "Recent test results of the high field Nb3Sn dipole magnet HD2", IEEE. Trans. Appl. Supercond. 20 (2010) 292.
[8] A. McInturff, et al., "Test results for a high field (13 T) Nb3Sn dipole", Particle Accelerator Conference (1997) 3212.
[9] B. Holzer, private communication.
[10] A. Lietzke, et al., "Test results for a $Nb_3Sn$ dipole magnet", IEEE Trans. Appl. Supercond. 7 (1997) 739.
[11] O. Bruning, et al, "LHC luminosity and energy upgrade: a feasibility study", LHC Project Report 626 (2002).
[12] P. McIntyre, A. Sattarov, "On the feasibility of a tripler upgrade for the LHC", PAC (2005) 634.
[13] R. Assmann, et al., "First thoughts on a higher energy LHC", CERN-ATS-2010-177 (2010).
[14] O. Bruning, et al., "LHC Design report", CERN 2004-003.
[15] S. Wolff, "The superconducting magnet system for HERA", proceedings of MT19, ed. By C. Marinucci and P. Waymuth (Zurich, SIN, 1995).
[16] C. Lorin, et al., "Predicting the quench behaviour of the LHC dipoles during commissioning", IEEE Trans. Appl. Supercond. 20 (2010) 135.
[17] H. Ten Kate, et al, "Critical current degradation in Nb3Sn cables under transverse pressure", IEEE Trans. Appl. Supercond. 3 (1993) 1334.
[18] P. Fessia, D. Perini, "A novel idea for coil collar structures in accelerator superconducting magnets", IEEE Trans. Appl. Supercond. 12 (2002) 202.
[19] B. Strauss, US-DOE program in HTS, private communication.
[20] R. Gupta, "Common coil magnet system for VLHC", Particle Accelerator Conference (1999), 3239.
[21] V. Kashikin, et. al, "Development and Test of Single-Layer Common Coil Dipole Wound With Reacted Nb3Sn Cable", IEEE Trans. Appl. Supercond. 14 (2004) 353.
[22] S. Russenschuck, "Field Computation for Accelerator Magnets" (Wiley, 2010)..
[23] The LHC study group, "Design study of the Large Hadron Collider", CERN 91-03 (1991).
[24] G. Velev, et al., "Field quality measurements and analysis of the LARP technology quadrupole models", IEEE Trans. Appl. Supercond. 18 (2008) 184.
[25] A. Milanese, CERN - Technology dept, private communication.
[26] F. Zimmermann, et al., these proceedings.